\documentclass[a4paper,11pt]{article}
\usepackage{pos}

\title{Charged-averaged elastic lepton-proton scattering cross section results from OLYMPUS}
\ShortTitle{Charge-averaged results from OLYMPUS}

\author*[a,b]{Axel Schmidt}

\affiliation[a]{Massachusetts Institute of Technology,\\
  77 Massachusetts Ave., Cambridge, MA 02139, USA}

\affiliation[b]{George Washington University,\\
  2121 I St.\ NW, Washington, DC 20052, USA}

\emailAdd{axelschmidt@gwu.edu}

\abstract{
Measurements of the proton's form factor ratio made with polarization transfer show a striking
discrepancy relative to the ratio extracted from unpolarized elastic electron-proton
scattering cross sections. One hypothesis is that the discrepancy is caused
by hard two-photon exchange (TPE), a typically neglected radiative correction that
may bias the two approaches differently. This hypothesis has been challenging to
confirm. Theoretical estimates of TPE are model-dependent, and recent experimental
determinations of TPE lacked the kinematic reach to be conclusive. The possible
impact of TPE remains a cloud over our knowledge of the proton's form factors.
Recently, the OLYMPUS experiment published new elastic scattering cross sections
that are insensitive to the effects of TPE: specifically the average of electron-proton
and positron-proton cross sections. The OLYMPUS experiment, conducted at DESY, Hamburg,
measured elastic $e^-p$ and $e^+p$ scattering by detecting the scattered lepton and recoiling
proton in coincidence in a large-acceptance, toroidal magnetic spectrometer.
OLYMPUS was designed to measure the $e^+p/e^-p$ cross section raio to isolate
the effects of TPE. By exploiting the over determined kinematics of the reaction,
the absolute efficiency of spectrometer could be verified, allowing cross sections
to be extracted from the data. These results can help refine our knowledge of the
proton's form factors, especially in the squared momentum-transfer region of
1--2 GeV$^2$, where some previous measurements are in tension.

}

\FullConference{%
  *** Particles and Nuclei International Conference - PANIC2021 ***\\
  *** 5 - 10 September, 2021 ***\\
  *** Online ***
}

\begin{document}
\maketitle

\section{Introduction}

The proton's electromagnetic form factors, $G_E$ and $G_M$, show a significant
discrepancy depending on the technique used to measure them.
When determined from unpolarized elastic electron scattering cross sections,
the form factors generally
exhibit scaling, i.e., the ratio $\mu_p G_E / G_M$ is approximately constant as
a function of momentum transfer, $Q^2$. By contrast, measurements of polarization
transfer or other equivalent polarization observables indicate a sharply decreasing
ratio, as shown in Fig.~\ref{fig:r2g}. The leading hypothesis is that the effects
of hard two-photon exchange (TPE), a typically neglected radiative correction, might be
affecting the two techniques differently~\cite{Guichon:2003qm,Blunden:2003sp}.
Several recent experiments have attemped to measure hard TPE
directly by comparing positron-proton and electron-proton elastic scattering
cross sections, but the results have not been conclusive~\cite{Rachek:2014fam,Adikaram:2014ykv,Rimal:2016tm,Henderson:2016dea}.
Accounting for hard TPE remains a problem for understanding the proton form factors at high $Q^2$.
For a recent review, see Ref.~\cite{Afanasev:2017gsk}. 

\begin{figure}[htpb]
  \centering
  \includegraphics{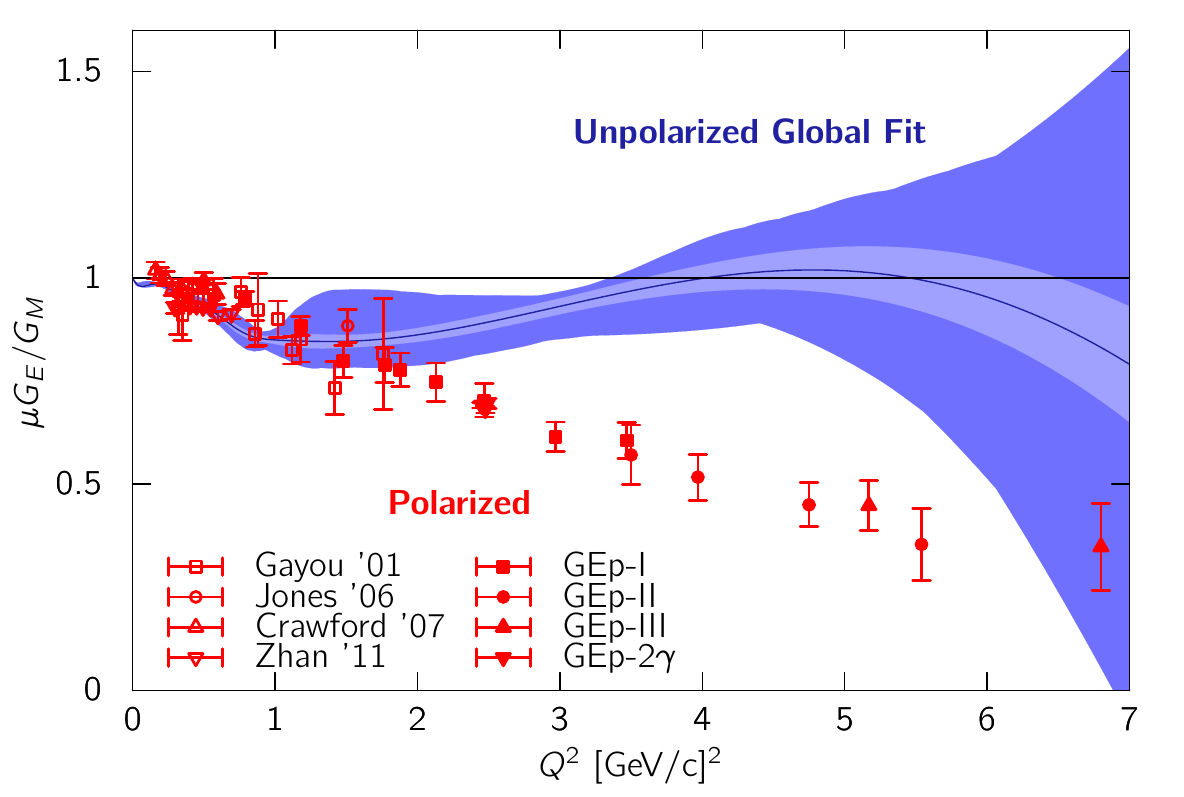}
  \caption{\label{fig:GE_over_GM}
    The protons form factor ratio, $\mu_p G_E/G_M$, shows a striking discrepancy between unpolarized Rosenbluth separations
    and direct measurements using polarization observables. The unpolarized global fit comes from Ref.~\cite{Bernauer:2013tpr}. The polarized
    data are from Refs.~\cite{Gayou:2001qt,Jones:2006kf,Crawford:2007dl,Zhan:2011ji,Punjabi:2005wq,Puckett:2011xg,Puckett:2017flj}.
  }
  \end{figure}

One of these recent experiments, OLYMPUS, has recently published absolute cross sections
for both electron and positron elastic scattering~\cite{OLYMPUS:2020dgl}. The average of
the two---the lepton charge averaged cross section---has the convenient feature of being
insensitive to two-photon exchange effects at lowest order. This can observable can help
constrain proton form factors by removing the uncertainty associated with two photon exchange
corrections.

\section{The OLYMPUS Experiment}

\begin{figure}[htpb]
  \centering
  \includegraphics[width=0.41\textwidth]{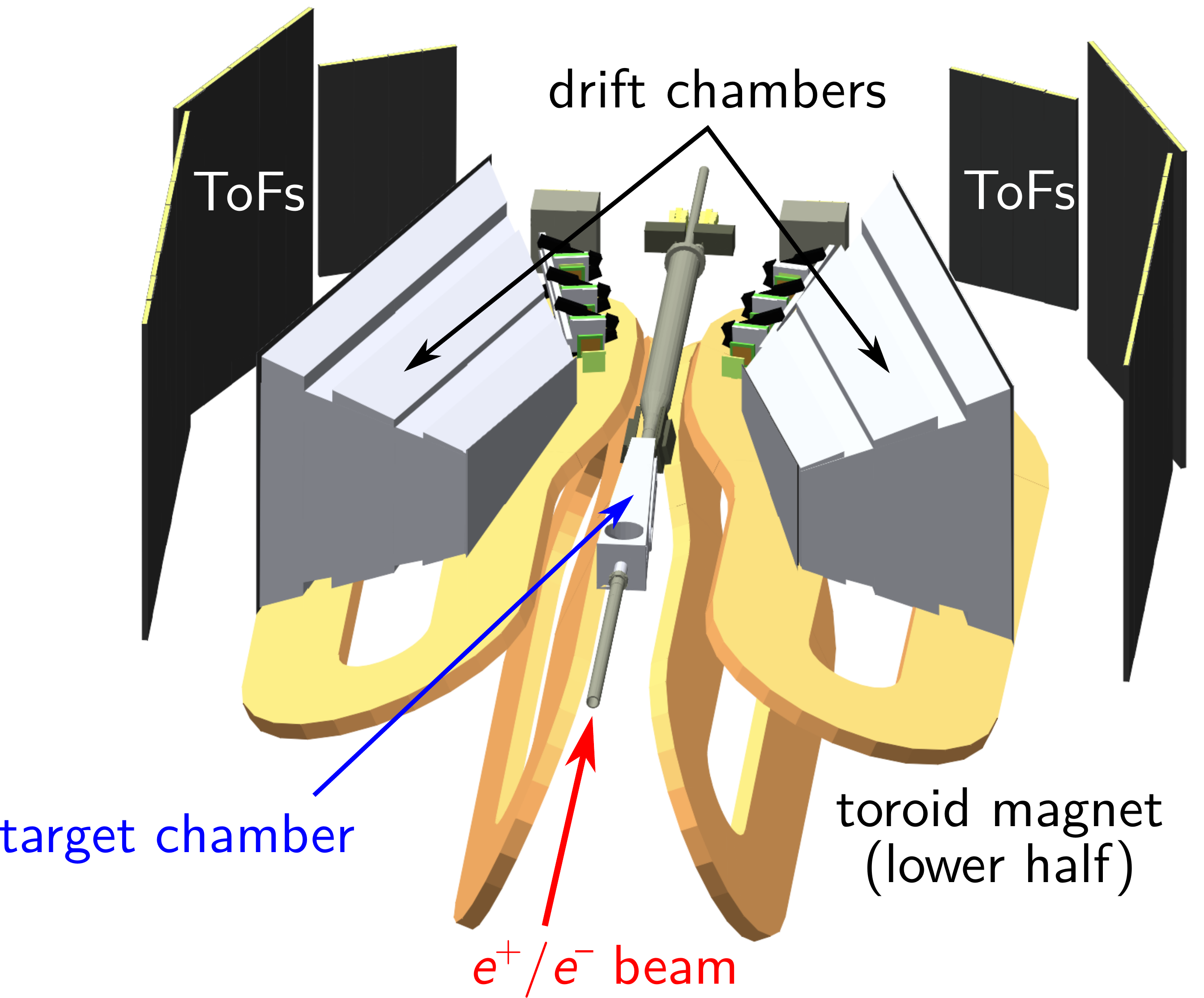}\hspace{0.03\textwidth}
  \includegraphics[width=0.49\textwidth]{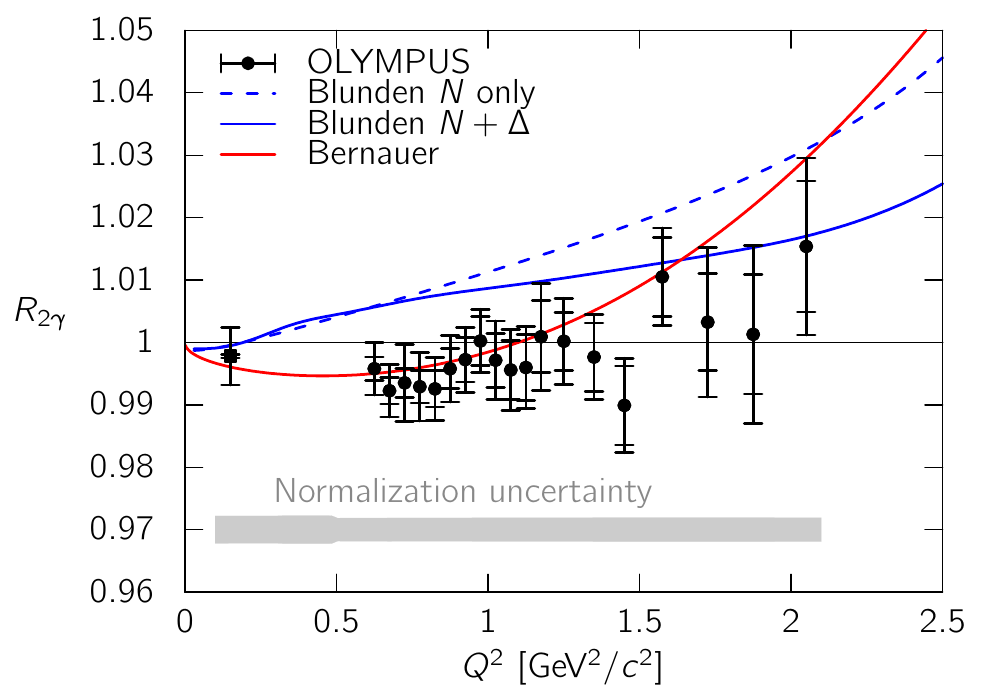}
  \caption{\label{fig:r2g}
    Left: illustration of the OLYMPUS detector, described in detail in Ref.~\cite{Milner:2013daa}.
    Right: the OLYMPUS measurement of $R_{2\gamma}$ \cite{Henderson:2016dea} showed a slight increase with $Q^2$ as
    expected from the two-photon exchange hypothesis, but was slightly lower than theoretical expectations~\cite{Blunden:2017nby,Bernauer:2013tpr}.
  }
  \end{figure}

The OLYMPUS Experiment collected data at DESY, in Hamburg, Germany in 2012. Alternating beams of electrons and positrons
in the DORIS storage ring were passed through a windowless unpolarized hydrogen target that was internal to the ring vacuum~\cite{Bernauer:2014pva}.
The OLYMPUS spectrometer~\cite{Milner:2013daa} consisted of two instrumented sectors of an 8-coil toroid magnet~\cite{Bernauer:2016cc} surrounding the
target, shown in Fig.~\ref{fig:r2g} left.
Drift chambers were used for charged particle tracking and panels time-of-flight (ToF) scintillators were used for
triggering, timing, and particle identification. Scattered leptons and recoiling protons were detected
in coincidence and over-determined kinematics were used to cleanly separate elastic scattering events from inelastic background.
The relative luminosity between electron and positron running modes was determined from the rate of multi-interaction events
in a pair of lead fluoride calorimeters positioned downstream at the angle of symmetric M\o ller/Bhabha scattering~\cite{Schmidt:2017jby}. 

OLYMPUS's primary result was a measurement of the ratio of positron-proton to electron-proton elastic scattering cross sections, i.e.,
$R_{2\gamma}\equiv \sigma_{e^+p}/\sigma_{e^-p}$, shown in Fig.~\ref{fig:r2g} right~\cite{Henderson:2016dea}. The results show a definitive
indication of TPE: $R_{2\gamma}$ has a positive slope with increasing $Q^2$ (and decreasing $\epsilon$).
However the results are generally below theoretical predictions~\cite{Blunden:2017nby,Bernauer:2013tpr}. It is not
clear whether this slight disagreement is due to additional effects not accounted for in the predictions or
to the overall normalization of the experiment (estimated to be better than $\pm 0.5\%$). In any case, 
the OLYMPUS results are completely consistent with the magnitude of the form factor discrepancy
in the $Q^2$ range covered by the experiment~\cite{Schmidt:2019vpr}.

\section{Charge-Averaged Cross Section Analysis}

While the ratio $\sigma_{e^+p}/\sigma_{e^-p}$ amplifies the effects of TPE, the
lepton charge-averaged cross section, $( \sigma_{e^+p} + \sigma_{e^-p})/2$, cancels the effects of
TPE at lowest order. This combination therefore provides a more accurate measure
of the Born cross section, removing the need to apply a correction for hard TPE.

Determining absolute cross sections was difficult for OLYMPUS, which was designed to measure
cross section ratios in which the effects of acceptance, efficiency, and absolute luminosity normalization
largely cancel. Nevertheless, the over-determined kinematics in elastic scattering allowed
several cross checks to be performed in order to quantify the uncertainty coming from each effect.

\begin{figure}[t]
  \centering
  \includegraphics{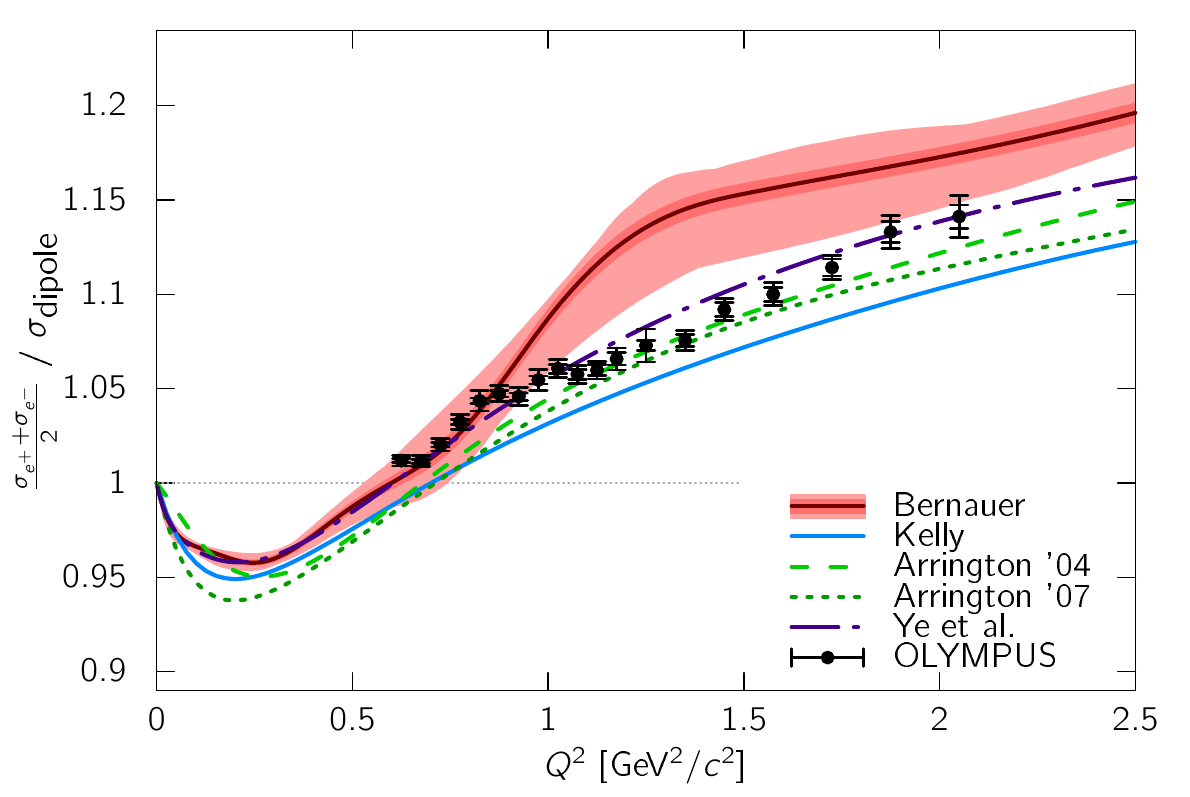}
  \caption{\label{fig:yield}
    The charge-averaged cross section as measured by OLYMPUS~\cite{OLYMPUS:2020dgl}, compared to other recent
    proton form-factor predictions~\cite{PhysRevC.70.068202,Arrington:2003qk,Arrington:2007ux,Ye:2017gyb}.
    A normalization uncertainty of $\pm 7.5\%$ is not shown.
  }
  \end{figure}

The efficiency of each sub-detector element---each scintillator panel and each drift chamber cell---was
determined from data and then modelled in a Geant4 simulation. The efficiency for reconstructing
elastic scattering events was determined from simulation. To cross check that there was no additional
tracking inefficiency in real data relative to simulated data, elastic events were identified using
the lepton (or proton) track information alone, to determine the probability for the successful
reconstruction of the opposite side proton (or lepton). There was no indication of any additional
tracking inefficiency, at the percent level.

OLYMPUS had no mechanism to deterimine the absolute luminosity, which is a combination of
the well-measured beam current and the crudely known density of the hydrogen gas target. 
The best measure of the absolute luminosity came from the forward M\o ller/Bhabha
calorimeters, but even this was estimated to have an accuracy of $\pm 7\%$. The
OLYMPUS results, therefore, are primarily a measure of the shape of the cross section,
rather than one of absolute scale. 

\section{Results}

The results~\cite{OLYMPUS:2020dgl} are shown in Fig.~\ref{fig:yield}, in comparison to other recent
global fits of proton cross section data~\cite{PhysRevC.70.068202,Arrington:2003qk,Arrington:2007ux,Ye:2017gyb}.
The results are in good qualitative agreement with these global fits. However, the data do not
show the cusp predicted by the Bernauer et al.\ fit~\cite{Bernauer:2013tpr}. It will be interesting
to see how this cusp changes when the Bernauer fit is updated to include the new OLYMPUS data.
In general, the OLYMPUS results can help constrain future global fits with reduced ambiguity from
uncertain corrections for hard TPE.

\section{Summary}

Data from the OLYMPUS Experiment were analyzed to extract absolute cross sections for both
electron-proton and positron-proton elastic scattering. The average of the two is unaffected
by hard TPE, giving robust access to the Born cross section and the proton's
electromagnetic form factors. These results will aid future global fits to constrain $G_E$
and $G_M$ in the momentum transfer range of $0.6<Q^2<2.0$~GeV$^2/c^2$.

\bibliographystyle{JHEP}
\bibliography{references}


\end{document}